# Superconductivity and Ferromagnetism from Effective Mass Reduction

J.E. Hirsch

Department of Physics, University of California, San Diego, La Jolla, CA 92093-0319, USA

Within a simple model Hamiltonian, both superconductivity and metallic ferromagnetism can be understood as arising from lowering of kinetic energy as the ordered state develops, due to a reduction in the carriers effective mass, or equivalently, a bandwidth expansion. Experimental manifestation of this physics has been detected in both high $T_c$ superconductors and large magnetoresistance ferromagnets, as an anomalous transfer of spectral weight in optical absorption from high to low frequencies as the ordered state develops. It is proposed that this general principle is common to the essential physics of superconductivity and ferromagnetism in nature, and hence that these effects in optical properties, although often smaller in magnitude, should exist in all superconductors and metallic ferromagnets.

Consider the low energy effective Hamiltonian describing interacting electrons in a band in a solid that crosses the Fermi level. In momentum space,

$$H = \sum_{k\sigma} \varepsilon_k c^+_{k\sigma} c_{k\sigma} + \sum_{kk'q} V_{kk'q} c^+_{k+q\sigma} c^+_{k'-q\sigma'} c_{k'\sigma'} c_{k\sigma} \quad (1)$$

where $c_{k\sigma}$ destroys an electron with Bloch wave function $\Psi_{nk}(r)$ in the n-th band in a solid (the band index is omitted in the operator since only one band will be considered). This Hamiltonian contains the band energy $\varepsilon_k$ obtained from solving the Hamiltonian for one electron interacting with the ionic potential, and the Coulomb term describing interactions between electrons in that band:

$$V_{kk'q} = \int d^3r d^3r' \, \Psi^*_{k+q}(r) \Psi^*_{k'-q}(r') \frac{e^2}{|r-r'|} \quad (2)$$
$$\times \Psi_{k'}(r') \Psi_k(r)$$

We ignore matrix elements of the Coulomb interaction that connect this band with other bands, under the assumption that they are unimportant for the physics of interest. This will be particularly accurate if other bands are well separated in energy, and the assumption will fail when there is band degeneracy. However, except for symmetry points or lines in the Brillouin zone, which are a set of measure zero, no degeneracies can exist for general points in the Brillouin zone.

Consider next a transformation from Bloch to Wannier states:

$$\varphi_{ni}(r) = \frac{1}{\sqrt{N}} \sum_k e^{ikR_i} \Psi_{nk}(r) \quad (3)$$

and let $c_{i\sigma}$ be the operator that destroys an electron in Wannier state $\varphi_{ni}(r)$. The Hamiltonian in this basis is (again we omit the band index from the operators for convenience):

$$H = -\sum_{<ij>\sigma} t_{ij}(c^+_{i\sigma}c_{i\sigma} + h.c.) + \quad (4a)$$

$$\sum_{ijkl\sigma\sigma'} (ij|1/r|kl) c^+_{i\sigma} c^+_{j\sigma'} c_{l\sigma'} c_{k\sigma}$$

$$(ij|1/r|kl) = \int d^3r d^3r' \, \varphi^*_{ni}(r) \varphi^*_{nj}(r') \times \quad (4b)$$
$$\frac{e^2}{|r-r'|} \varphi_{nl}(r') \varphi_{nk}(r)$$

and $t_{ij}$ the matrix elements of the single particle part of the Hamiltonian with the Wannier states. There is five types of interaction terms involving one and two-center integrals[1,2]: on-site (U) and nearest neighbor (V) repulsion, correlated hopping ($\Delta t$), exchange (J), and pair hopping (J'). The simplest Hamiltonian that can describe superconductivity and ferromagnetism has interactions U, J and $\Delta t$ only[2]:

$$H = -\sum_{<ij>\sigma} t_{ij}(c^+_{i\sigma}c_{i\sigma} + h.c.) +$$

$$\Delta t \sum_{<ij>\sigma} (n_{i,-\sigma} + n_{j,-\sigma})(c^+_{i\sigma}c_{j\sigma} + h.c.) + \quad (5)$$

$$U\sum_i n_{i\uparrow} n_{i\downarrow} + J \sum_{<ij>\sigma\sigma'} c^+_{i\sigma} c^+_{j\sigma'} c_{i\sigma'} c_{j\sigma}$$

where sites <ij> are nearest neighbors. We have omitted other interaction terms, some of which are larger in magnitude that those included in Eq. (5), such as the nearest neighbor repulsion V, as well as all terms involving further than nearest neighbors. Such terms may be quantitatively important but we do not expect them to change the qualitative physics.

Within the Hamiltonian Eq. (5), the term involving $\Delta t$ gives rise to superconductivity[3], and the one involving J gives rise to ferromagnetism[4]. U plays an important role in suppressing superconductivity and enhancing ferromagnetism. As discussed in Refs. 2-4, both $\Delta t$ and J cause enhancement of the carriers effective mass in the normal state. Pairing (of holes) leads to a local lowering of the electronic site charge, and this leads to an enhanced ability for the electrons (or holes) to hop, described in Eq. (5) by the $\Delta t$ term. Similarly, spin polarization leads to a local lowering of the electronic bond charge, and this leads to an enhanced ability for the carriers to hop described by the J term in Eq. (5). Hence these interactions cause a lowering of the carriers effective mass, and kinetic energy, as the transition to the ordered state occurs.

At the time that the ideas discussed here were first proposed[2-4] there was no experimental evidence in any material that either superconductivity or ferromagnetism might be driven by lowering of kinetic energy. Since then, such experimental evidence has been detected. In superconductors, optical experiments by Basov et al[5] probing c-axis transport have detected an apparent violation of the Ferrell-Glover-Tinkham sum rule[6], indicating lowering of kinetic energy as the system enters the superconducting state. This phenomenon had been discussed in connection with the model discussed here for high $T_c$ oxides in 1992[7]. The apparent disagreement with experiment, that detects a large c-axis sum rule violation and has so far not seen any in-plane effect of this sort[5], is explained[8] by the fact that high $T_c$ cuprates appear to be in the clean limit for in-plane transport and in the dirty limit for c-axis transport[9], which greatly enhances the effect in the c direction.

In ferromagnets, clear evidence for effective mass reduction, or bandwidth expansion, as ferromagnetism develops, has been now seen in optical properties of colossal magnetoresistance manganites[10] and of some rare earth hexaborides[11]. There is however another signature of effective mass reduction due to ferromagnetism: negative magnetoresistance, which is universally seen in metallic ferromagnets. This is a consequence of the model Eq. (5), which can describe both regimes where the magnitude of the magnetoresistance is a few percent (e.g. weak ferromagnets, Fe, etc.[4]) and where it is very large as in the manganites and hexaborides[12]. Consequences of this model for optical properties as function of temperature and magnetic field are also discussed in Ref. 12.

If kinetic energy lowering occurs in the cases mentioned above, it is possible that it occurs in all transitions to superconductivity and ferromagnetism, as predicted by the model Hamiltonian Eq. (5), even though in many cases it may be difficult to detect. If so, the condensation energy in superconductors would arise from kinetic energy rather than potential energy, as is assumed in most other models such as electron-phonon. Similarly, in ferromagnets the energy lowering would be due to kinetic energy rather than exchange energy as usually assumed. It is natural to expect that for metals, defined by their ability to conduct electricity, transitions to collective states would be driven by the goal of becoming better conductors through effective mass reduction.